\documentstyle[prd,aps]{revtex}

\newcommand{\bea}{\begin{eqnarray}}
\newcommand{\eea}{\end{eqnarray}}

\begin{document}
\draft
\twocolumn[\hsize\textwidth\columnwidth\hsize\csname
@twocolumnfalse\endcsname

\title{A Conserved variable in the perturbed hydrodynamic world model}
\author{Jai-chan Hwang}
\address{Department of Astronomy and Atmospheric Sciences,
         Kyungpook National University, Taegu, Korea}

\date{\today}
\maketitle

\begin{abstract}

We introduce a scalar-type perturbation variable $\Phi$ which is conserved
in the large-scale limit considering general sign of three-space 
curvature ($K$), the cosmological constant ($\Lambda$), and 
time varying equation of state.
In a pressureless medium $\Phi$ is {\it exactly conserved} in all scales.

\end{abstract}

\vskip2pc]

{\it 1. Introduction:}
Relativistic cosmological perturbations with hydrodynamic energy-momentum
tensor was originally studied by Lifshitz in 1946 \cite{Lifshitz}
based on the synchronous gauge.
More convenient analyses based on better suited gauge conditions
were made by Harrison in 1967 \cite{Harrison} and Nariai in 1969 \cite{Nariai}.
Although the variables used by Harrison and Nariai are free of gauge
degree of freedom (thus, equivalent to corresponding gauge-invariant variables),
these gauge conditions became widely known by a seminal paper by Bardeen 
in 1980 \cite{Bardeen}.
We believe these are brighter side of the history concerning gauge condition
in the cosmological perturbation: Lifshitz carefully traced the remaining
gauge solutions, Harrison and Nariai in fact hit the correct gauge
conditions for handling aspects of hydrodynamic perturbations
[see below Eq. (\ref{varphi-chi-sol})], and Bardeen showed the gauge 
invariance of 
the variables used by Harrison and Nariai and demonstrated the diversity 
of gauge conditions and gauge-invariant way of handling them.
On the other side, there exist persistent {algebraic errors} in the 
literature which are often claimed to be due to wrong gauge conditions
\cite{SG-errors}.
These errors probably gave some researchers an (inappropriate) impression 
that the field is `plagued with gauge problems'.
In any case, there exist {\it many} gauge conditions waiting to be employed
with possible advantages in exploring certain aspects of problems.
In 1988 Bardeen \cite{Bardeen-1988} made a practical suggestion  
concerning the gauge condition which allows maximal use of the
various different gauge conditions depending on problems.
In \cite{PRW} we elaborated the suggestion and recently termed our approach
a {\it gauge-ready mothod}; see Sec. {\it 2}.

We may not need to emphasize the importance of conserved quantities in
physical processes.
In cosmological perturbations the conserved variable can provide
easy connection between the final results and the initial conditions.
Aspects of conserved perturbation variables were discussed in 
\cite{BST,Lyth,Conservation,Dunsby-Bruni,Zimdahl}.
In this paper, based on the gauge-ready method, we will 
derive a scalar-type perturbation variable which is 
conserved in the large-scale limit independently of changing
background world model with general $K$, $\Lambda$, and the 
perfect fluid equation of state.

Besides the scalar-type perturbation we also have the vector-type (rotation)
and the tensor-type (gravitational wave) perturbations which evolve
independently to the linear order in our simple background world model.
We also have conserved quantities for these additional perturbations: 
ignoring anisotropic stresses, 
the angular-momentum of the rotation is generally conserved,
whereas the non-transient solution of the graviational wave is 
conserved in the super-horizon scale in near flat case.
These two conservation properties were already noticed in \cite{Lifshitz},
and recently, we have elaborated these conservation properties in some
general situations \cite{GGT-conserv}.
In the following we concentrate on the scalar-type perturbation
with the hydrodynamic energy-momentum tensor.

Although redundant, in order to make this paper self-contained, 
in the Appendix we present the complete set of perturbed equations 
based on Einstein equations.
Our new result is the conserved variable in Eq. (\ref{Phi-def}) with the 
equation and the large-scale solution in Eqs. (\ref{Phi-eq},\ref{Phi-sol}).

\vskip .5cm
{\it 2. Notations and strategy:}
We consider {\it the most general} scalar-type perturbations in the 
spatially homogeneous and isotropic world model.
Our notations for the metric and the energy-momentum tensor are:
\bea
   & & d s^2 = - a^2 \left( 1 + 2 \alpha \right) d \eta^2
       - a^2 \beta_{,\alpha} d \eta d x^\alpha
   \nonumber \\
   & & \qquad
       + a^2 \left[ g_{\alpha\beta}^{(3)} \left( 1 + 2 \varphi \right)
       + 2 \nabla_\alpha \nabla_\beta \gamma \right] d x^\alpha d x^\beta,
   \label{metric} \\
   & & T^0_0 = - \left( \bar \mu + \delta \mu \right), \quad
       T^0_\alpha = - {1 \over k} \left( \mu + p \right) v_{,\alpha}, 
   \nonumber \\
   & & T^\alpha_\beta
       = \left( \bar p + \delta p \right) \delta^\alpha_\beta
       + {1 \over a^2} \left( \nabla^\alpha \nabla_\beta
       - {1 \over 3} \Delta \delta^\alpha_\beta \right) \sigma,
   \label{Tab}
\eea
where $0 = \eta$, and an overbar indicates a 
background order quantity and will be ignored unless necessary.
Spatial indices ($\alpha$, $\beta, \dots$) and $\nabla_\alpha$ are based on 
$g^{(3)}_{\alpha\beta}$
which is the three-space metric of the homogeneous and isotropic space.
$\beta$ and $\gamma$ always appear in a spatially gauge-invariant
combination $\chi \equiv a ( \beta + a \dot \gamma)$; 
an overdot denotes the time derivative based on $t$ with $dt \equiv a d \eta$.
Using $\chi$, all the perturbed metric and energy-momentum tensor variables
in Eqs. (\ref{metric},\ref{Tab}) are spatially gauge-invariant.

Perturbed order variables can be expanded in eigenfunctions of the
Laplace-Beltrami operator ($\Delta$ based on $g^{(3)}_{\alpha\beta}$) 
with eigenvalues $-k^2$ where $k$ is a comoving wave number \cite{Bardeen};
to the linear order each eigenmode decouples, and without any confusion we can
assume variable in either the configuration space or the phase space.
{}For the flat ($K= 0$) and the hyperbolic ($K = -1$) backgrounds
$k^2$ takes continuous value with $k^2 \ge 0$, whereas in the spherical
($K = +1$) background we have $k^2 = n^2 - K$ $(n = 1,2,3,\dots)$
\cite{Lifshitz,Harrison,spherical,Lyth-Woszczyna}.
Situation in the hyperbolic background may deserve special attention.
Probably because any square integrable function can be expanded using harmonics
with $k^2 \ge 1$ (subcurvature) modes only, most of the cosmology literature
ignored $0\le k^2 <1 $ (supercurvature) modes, and gave a wrong impression
that the supercurvature modes do not exist 
\cite{Lifshitz,Harrison}: 
a state of affairs was well summarized in \cite{Lyth-Woszczyna}.
In the spherical background, the lack of physically relevant perturbations
for the lowest two harmonics $n = 1$ ($k^2 = 0$) and
$n = 2$ ($k^2 - 3 K = 0$) was pointed out in \cite{Lifshitz}.
Later, we will use vanishing $c_s^2 k^2$ term as the large-scale limit.
Thus the results in such a limit will be relevant in
all scales for general $K$ in the pressureless medium ($c_s^2 = 0$),
and in the large-scale limit ($k^2 \rightarrow 0$) for the flat and 
the hyperbolic situations in a medium with $c_s^2 \sim 1 (\equiv c^2)$.

Complete sets of equations for the background and the perturbed orders
are presented in the Appendix.
Equations (\ref{eq1}-\ref{eq7}) are written in a gauge-ready form.
In this form we still have a {\it right} to choose a temporal gauge 
condition, \cite{PRW}.
Equations are designed so that imposing the gauge condition is as 
simple as the following: the synchronous gauge chooses $\alpha \equiv 0$,
the comoving gauge chooses $v/k \equiv 0$, 
the zero-shear gauge chooses $\chi \equiv 0$,
the uniform-curvature gauge chooses $\varphi \equiv 0$,
the uniform-expansion gauge chooses $\kappa \equiv 0$, 
the uniform-density gauge chooses $\delta \mu \equiv 0$, etc. 
Except for the synchronous gauge,
since any of the other gauge conditions completely fixes the temporal gauge 
degree of freedom [see Eq. (\ref{GT})], any variable in such a gauge
condition is equivalent to a unique gauge-invariant combination of
the variable concerned and the variable used in the gauge condition.
Examples of some useful gauge-invariant combinations can be constructed
using Eq. (\ref{GT}):
\bea
   & & \varphi_v \equiv \varphi - {a H \over k} v, \quad
       \varphi_\chi \equiv \varphi - H \chi, 
   \nonumber \\
   & & \delta_v \equiv \delta + 3 (1 + {\rm w}) {aH \over k} v, \quad
       v_\chi \equiv v - {k \over a} \chi.
   \label{GI}
\eea
In this way, we can flexibly use the gauge degree of freedom as an 
{\it advantage} in handling problems \cite{Bardeen-1988,PRW}.

\vskip .5cm
{\it 3. A conserved variable:}
{}From Eqs. (\ref{eq1},\ref{eq2},\ref{eq4},\ref{eq5}) we can derive
\cite{Derivation-comment}
\bea
   & & {\mu + p \over H} \left[ {H^2 \over ( \mu + p) a}
       \left( {a \over H} \varphi_\chi \right)^\cdot \right]^\cdot
       + c_s^2 {k^2 \over a^2} \varphi_\chi = {\rm stresses}.
   \label{varphi-chi-eq}
\eea
In the large-scale limit, super-sound-horizon scale where we ignore 
$c_s^2 k^2/a^2$ term \cite{comment:large-scale}, and ignoring stresses 
($e$ and $\sigma$), 
Eq. (\ref{varphi-chi-eq}) has an exact solution 
valid for general $K$, $\Lambda$ and time-varying equation of state
$p(\mu)$, \cite{HV,Newtonian}
\bea
   \varphi_\chi ({\bf x}, t)
       = C ({\bf x}) 4 \pi G {H \over a} \int_0^t {(\mu + p) a \over H^2} dt
       + d ({\bf x}) {H \over a}, 
   \label{varphi-chi-sol}
\eea
where $C ({\bf x})$ and $d ({\bf x})$ are coefficients of the growing
and decaying solutions, respectively.
$\varphi_\chi$ most closely resembles the behavior of perturbed
Newtonian potential \cite{Newtonian}.
The variables most closely resembling the Newtonian behaviors of 
the density and velocity perturbations are $\delta_v$ and $v_\chi$,
respectively \cite{Harrison,Nariai,Bardeen,Newtonian}.
{}From Eqs. (\ref{eq2},\ref{eq3}), and Eqs. (\ref{eq1},\ref{eq3},\ref{eq4}), 
we can derive, respectively:
\bea
   & & {k^2 - 3 K \over a^2} \varphi_\chi = 4 \pi G \mu \delta_v,
   \label{Poisson-eq} \\
   & & \dot \varphi_\chi + H \varphi_\chi
       = - 4 \pi G \left( \mu + p \right) {a \over k} v_\chi
       - 8 \pi G H \sigma.
   \label{velocity-eq}
\eea
Solutions for $\delta_v$ and $v_\chi$ follow from these equations.
We can similarly derive a solution for $\varphi_v$ using
\bea
   \varphi_v \equiv \varphi - {aH \over k} v 
       = \varphi_\chi - {aH \over k} v_\chi.
   \label{varphi-v} 
\eea

Introduce 
\bea
   \Phi \equiv \varphi_v - {K /a^2 \over 4 \pi G ( \mu + p) } \varphi_\chi
       = \varphi_v - {K \over k^2 - 3 K} {\delta_v \over 1 + {\rm w}},
   \label{Phi-def}
\eea
where we used Eq. (\ref{Poisson-eq}), \cite{Dunsby-comment}.
Using Eqs. (\ref{varphi-v},\ref{velocity-eq}), ignoring $\sigma$, we can show
\bea
   \Phi = {H^2 \over 4 \pi G ( \mu + p) a} 
       \left( {a \over H} \varphi_\chi \right)^\cdot. 
   \label{Phi-2}
\eea
Thus, using the large-scale solution in Eq. (\ref{varphi-chi-sol}) we have
\bea
   \Phi ({\bf x}, t) = C ({\bf x}),
   \label{Phi-sol}
\eea
where the decaying solution has vanished.
Therefore, {\it $\Phi$ is generaly conserved in the super-sound-horizon scale
considering general $K$, $\Lambda$ and time-varying $p (\mu)$.}
On the other hand, from Eqs. (\ref{varphi-chi-eq},\ref{Phi-2}),
thus ignoring stresses, we have
\bea
   \dot \Phi = - {H c_s^2 \over 4 \pi G ( \mu + p) } {k^2 \over a^2} 
       \varphi_\chi.
   \label{Phi-dot-eq}
\eea
Thus, {\it in a pressureless case $\Phi ({\bf x}, t) = C ({\bf x})$
is an exact solution valid in general scale};
this was noticed in \cite{Dunsby-Bruni}.
Combining Eqs. (\ref{Phi-2},\ref{Phi-dot-eq}) we have a closed form equation
for $\Phi$ 
\bea
   {H^2 c_s^2 \over ( \mu + p) a^3} 
       \left[ {( \mu + p) a^3 \over H^2 c_s^2} \dot \Phi \right]^\cdot
       + c_s^2 {k^2 \over a^2} \Phi = 0,
   \label{Phi-eq} 
\eea
which is valid for $c_s^2 \neq 0$.
Thus, for vanishing $c_s^2 k^2/a^2$ term \cite{comment:large-scale}
we have a general solution
\bea
   \Phi ({\bf x}, t) = C ({\bf x}) + \tilde d ({\bf x}) \int_0^t {H^2 c_s^2 
       \over 4 \pi G ( \mu + p) a^3} dt,
   \label{Phi-sol2}
\eea
which includes $c_s^2 = 0$ limit.
We can show easily that $\tilde d$ term in Eq. (\ref{Phi-sol2})
is higher-order in the large-scale expansion compared with $d$ term 
in Eq. (\ref{varphi-chi-sol}):
by comparing the two solutions of Eqs. (\ref{varphi-chi-sol},\ref{Phi-sol2})
in Eq. (\ref{Phi-dot-eq}) we can show $\tilde d = - k^2 d = \Delta d$.
Therefore, the solution in Eq. (\ref{Phi-sol}) is valid in the 
large-scale (super-sound-horizon scale) with vanishing dominating
decaying solution.

\vskip .5cm
{\it 4. Other conservation variables:}
$\Phi$ differs from the well known conserved variable $\zeta$ 
in \cite{BST,Bardeen-1988}.
In \cite{BST} $\zeta$ was introduced in the flat background and in that
background it is the same as
\bea
   \zeta = \varphi + {\delta \over 3 (1 + {\rm w})} \equiv \varphi_\delta,
   \label{zeta}
\eea
whereas, in the flat background we have $\Phi = \varphi_v$
\cite{comment:zeta}.
It is interesting to note that $\varphi_\delta$ is also conserved
in the large-scale limit considering general $K$ \cite{Newtonian}:
from Eqs. (\ref{Poisson-eq},\ref{Phi-def},\ref{zeta}), we can derive
\bea
    \varphi_\delta = \Phi + {1 \over 12 \pi G ( \mu + p)}
        {k^2 \over a^2} \varphi_\chi.
    \label{varphi-delta-eq}
\eea
According to the solutions in Eqs. (\ref{varphi-chi-sol},\ref{Phi-sol}),
or Eq. (\ref{Phi-sol2}), we can see that for vanishing $k^2$ order term
$\varphi_\delta$ is conserved considering general $K$.
However, we also see that due to the second term the conservation property 
of $\varphi_\delta$ breaks down near and inside horizon ($k/[aH] \ge 1$) 
even for $K = 0$; whereas, $\Phi$ is conserved independently of the 
horizon crossing in the matter dominated era.
We can also show the conservation property of $\varphi_\kappa$:
from Eqs. (\ref{GT},\ref{eq2}) we can show
\bea
   \varphi_\kappa \equiv \varphi + {H \kappa \over 3 \dot H - k^2/a^2}
       = {\varphi_\delta \over 1 
       + {k^2 - 3 K \over 12 \pi G ( \mu + p) a^2} }.
   \label{varphi-kappa-sol}
\eea
Thus, the conservation property of $\varphi_\kappa$ breaks down
for $K \neq 0$ or near and inside horizon.
{}In the $K = 0$ situation, $\varphi$ in many different gauge conditions shows 
the conserved behavior in the large-scale limit: for the ideal fluid see 
Eqs. (41,73) in \cite{Ideal}, and Eq. (34,35) in \cite{MDE};
for the scalar field see Eqs. (92) in \cite{MSF};
and for the generalized gravity see Sec. VI in \cite{GGT-HN}.
Conservation properties of $\varphi$ in various gauge conditions
were also discussed in \cite{Dunsby-Bruni,Zimdahl}
where the arguments were based on the first order equations of the type
in Eq. (\ref{Phi-dot-eq}).

\vskip .5cm
{\it 5. Discussions:}
We would like to emphasize again that the conservation property of
$\Phi$ is valid in the limit of vanishing $c_s^2 k^2 /a^2$ term.
Thus, in the pressureless medium ($c_s^2 = 0$) it applies {\it in all scales 
for general $K$},
and in the medium with dominant pressure ($c_s^2 \sim 1$) 
it applies {\it in the large-scale limit $(k^2 \rightarrow 0)$ for 
the flat and the hyperbolic situation}.
As long as these conditions are met, $\Phi$ is conserved independently
of time varying equation of state $p(\mu)$.
Since we anticipate time varying equation of states during equal-time 
from the radiation dominated era ($p = {1 \over 3} \mu$) to 
the matter dominated era ($p = 0$), and during (p)reheating period
from the acceleration era ($p<-{1 \over 3} \mu$) to the radiation era,
and since the observationally relevant scales stayed in the large-scale limit
during the transitions,
$\Phi$ is a practically important quantity in tracing the evolution of
scalar-type perturbation from the early universe till recent era
before the nonlinear evolution takes over.

We would like to conclude by remarks on the history about the variables
in Eq. (\ref{Phi-def}):
$\varphi_\chi$ and $\delta_v$ in their gauge-invariant forms
became widely known by Bardeen's work in 1980 
(these are $\Phi_H$ and $\epsilon_m$ in \cite{Bardeen}).
However, $\varphi_\chi$ is the same as $\varphi$ in the zero-shear gauge
[which fixes $\chi = 0$ as the temporal gauge condition, 
see Eq. (\ref{GI})] which was first used by Harrison in 1967,
and $\delta_v$ is the same as $\delta$ in the comoving gauge
(which fixes $v/k = 0$) which was first used by Nariai in 1969 \cite{Nariai}.
{}For $K = 0$, the variable $\varphi_v$ is widely recognized as a 
conserved variable in the literature.
Up to our knowledge it was first introduced as $\varphi$ in the
comoving gauge by Lyth in 1985 \cite{Lyth},
and later was used in the context of the scalar field and generalized gravity
as the large-scale conserved variable \cite{varphi-v}.

\vskip .5cm
We thank Peter Dunsby, Hyerim Noh and Winfried Zimdahl for many useful 
discussions.
We wish to thank P. Dunsby for encouraging us to find the conservation 
variable for general $K$ and for hospitality during 
our visit University of Cape Town where the result was found.

\vskip .5cm
\centerline{\bf Appendix}

In the following we {\it derive} Einstein equations based on 
Eqs. (\ref{metric},\ref{Tab}).
To the background order, from $T^0_0$ and $T^\alpha_\alpha - 3 T^0_0$
components of Einstein equations and $T^b_{0;b} = 0$, respectively,
we can derive:
\bea
   & & H^2 = {8 \pi G \over 3} \mu - {K \over a^2} + {\Lambda \over 3}, \quad
       \dot H = - 4 \pi G ( \mu + p ) + {K \over a^2}, \quad
   \nonumber \\
   & & \dot \mu = - 3 H \left( \mu + p \right),
   \label{BG-eqs}
\eea
where $H \equiv {\dot a \over a}$.
To the perturbed order we can derive:
\bea
   & & \kappa \equiv 3 \left( - \dot \varphi + H \alpha \right)
       + {k^2 \over a^2} \chi,
   \label{eq1} \\
   & & - {k^2 - 3K \over a^2} \varphi + H \kappa = - 4 \pi G \mu \delta,
   \label{eq2} \\
   & & \kappa - {k^2 - 3K \over a^2} \chi = 12 \pi G \left( \mu + p \right)
       {a \over k} v,
   \label{eq3} \\
   & & \dot \chi + H \chi - \alpha - \varphi = 8 \pi G \sigma,
   \label{eq4} \\
   & & \dot \kappa + 2 H \kappa = \left( {k^2 \over a^2} - 3 \dot H \right)
       \alpha + 4 \pi G \left( 1 + 3 c_s^2 \right) \mu \delta
   \nonumber \\
   & & \qquad
       + 12 \pi G e,
   \label{eq5} \\
   & & \dot \delta + 3 H \left( c_s^2 - {\rm w} \right) \delta
       + 3 H { e \over \mu } 
   \nonumber \\
   & & \qquad
       = \left( 1 + {\rm w} \right)
       \left( \kappa - 3 H \alpha - {k \over a} v \right),
   \label{eq6} \\
   & & \dot v + \left( 1 - 3 c_s^2 \right) H v = {k \over a} \alpha
   \nonumber \\
   & & \qquad
       + {k \over a \left( 1 + {\rm w} \right) } \left( c_s^2 \delta
       + {e \over \mu} - {2\over 3} {k^2 - 3K \over a^2} {\sigma \over \mu}
       \right),
   \label{eq7}
\eea
where we have introduced
\bea
   & & \delta p ({\bf k}, t) \equiv c_s^2 (t) \delta \mu ({\bf k}, t)
       + e ({\bf k}, t), 
   \nonumber \\
   & & \delta \equiv {\delta \mu \over \mu}, \quad
       {\rm w} (t) \equiv {p \over \mu}, \quad
       c_s^2 (t) \equiv {\dot p \over \dot \mu}.
   \label{definitions}
\eea
In Eq. (\ref{eq1}) we introduced a variable $\kappa$ which is the
perturbed part of the trace of extrinsic curvature; similarly, for meanings of 
the other perturbation variables, see Sec. 2.1 of \cite{PRW}.
Equations (\ref{eq2}-\ref{eq5}) follow from $T^0_0$, $T^0_\alpha$, 
$T^\alpha_\beta - {1 \over 3} \delta^\alpha_\beta T^\gamma_\gamma$,
and $T^\alpha_\alpha - T^0_0$ components of Einstein equations, respectively;
and Eqs. (\ref{eq6},\ref{eq7}) follow from $T^b_{0;b} = 0$
and $T^b_{\alpha ;b} = 0$, respectively.
This set of equations were originally derived in Eqs. (41-47) of 
\cite{Bardeen-1988}, see also Eqs. (22-28) in \cite{PRW}.

Under the gauge transformation $\tilde x^a = x^a + \xi^a$, we have
(see Sec. 2.2 in \cite{PRW}):
\bea
   & & \tilde \alpha = \alpha - \dot \xi^t, \;\;
       \tilde \varphi = \varphi - H \xi^t, \;\;
       \tilde \chi = \chi - \xi^t, \;\;
       \tilde v = v - {k \over a} \xi^t, 
   \nonumber \\
   & & \tilde \kappa = \kappa
       + \left( 3 \dot H - {k^2 \over a^2} \right) \xi^t, \quad
       \tilde \delta = \delta + 3 \left( 1 + {\rm w} \right) H \xi^t.
   \label{GT}
\eea



\begin{thebibliography}{99}
\bibitem{SG-errors}
         We found a common error in several textbooks \cite{textbooks} based on 
         the synchronous gauge:
         Eq. (15.10.57) in Weinberg 
         Eq. (10.118) in Peebles,
         Eq. (11.5.2) in Coles and Lucchin, 
         Eq. (8.52) in Moss, and
         problem 6.10 in Padmanabhan are in errors. 
         Notice that these are basically algebraic errors partly due to
         unnecessarily complicated algebra in the synchronous gauge,
         see \cite{Newtonian}.
\bibitem{Derivation-comment}
         In order to derive the equation for $\varphi_\chi$ 
         using Eqs. (\ref{eq1}-\ref{eq7}) 
         we simply set $\chi = 0$ in the equations, and then each 
         variable is equivalent to a gauge invariant combination
         with subindex $\chi$.
         Remember that $\varphi_\chi$ is the same as $\varphi$ in the
         zero-shear gauge which sets $\chi = 0$ as the gauge condition.
\bibitem{comment:large-scale}
         We would like to comment on our large-scale limit used in 
	 Eqs. (\ref{varphi-chi-eq},\ref{Phi-eq}).
	 Consider an equation of a form
	 \bea
	     {1 \over Z} ( Z \dot \phi )^\cdot
	         + c_s^2 {k^2 \over a^2} \phi = 0.
             \label{phi-eq}
         \eea
	 In terms of $\psi \equiv \sqrt{Z/a} \phi$ we have 
         ($\prime = {\partial \over \partial \eta}$)
	 \bea
	     \psi^{\prime\prime} + \left[ c_s^2 k^2
	         - {(\sqrt{Z/a})^{\prime\prime} / \sqrt{Z/a}} \right] \psi = 0.
             \label{psi-eq}
         \eea
	 By ignoring $c_s^2 k^2$ term in either equation we have
	 \bea
	     \phi \propto {\rm constant}, \; \int_0^t {dt / Z}.
         \eea
	 Thus, when we ignore $c_s^2 k^2 /a^2$ term in Eq. (\ref{phi-eq})
	 as the large-scale limit, we assume the second term in
	 Eq. (\ref{psi-eq}) is negligible compared with the third term.
\bibitem{Dunsby-comment}
         After completion of a draft P. Dunsby informed us that $\Phi$ was 
         also introduced in \cite{Dunsby-Bruni} in the covariant formulation, 
         see Eq. (16) in \cite{Dunsby-Bruni}.
         We notice that authors of \cite{Dunsby-Bruni} took
         $k^2 \rightarrow 1$ as the large-scale limit in the hyperbolic
         case, thus leading to different conclusions from ours: their 
         conclusion was that, except for the pressureless case, the 
         conservation property of $\Phi$ is valid only in near flat situation.
\bibitem{comment:zeta}
         Since $\zeta$ is gauge invariant we can evaluate Eq. (\ref{zeta}) 
         in the comoving gauge, thus show 
         $\zeta = \varphi_v + \delta_v/(3 + 3 {\rm w}) \neq \varphi_v$.
\end{thebibliography}
\end{document}